\documentclass[a4paper]{jpconf}
\usepackage{graphicx}

\usepackage{xspace}
\usepackage{paralist}

\newcommand{\apenetp}{APEnet+\xspace}
\newcommand{\apelink}{APElink\xspace}
\newcommand{\nanet}{NaNet\xspace}
\newcommand{\nanetone}{NaNet-1\xspace}
\newcommand{\pcie}{PCIe\xspace}
\newcommand{\realtime}{real-time\xspace}
\newcommand{\nvidia}{NVIDIA\xspace}
\newcommand{\nios}{\texttt{Nios~II}\xspace}

\newcommand{\ie}{\textit{i.e.}\xspace}
\newcommand{\gbe}{GbE\xspace}
\newcommand{\tengbe}{10GbE\xspace}
\newcommand{\eg}{\textit{e.g.}\xspace}

\begin{document}
\title{\nanet: a low-latency NIC enabling \mbox{GPU-based},
  \mbox{real-time} low level trigger systems.}


\author{Roberto Ammendola$^1$, Andrea Biagioni$^2$, Riccardo Fantechi$^3$ $^4$,
  Ottorino Frezza$^2$, Gianluca Lamanna$^4$, Francesca Lo Cicero$^2$, Alessandro Lonardo$^2$,
  Pier Stanislao Paolucci$^2$, Felice Pantaleo$^5$ $^4$, Roberto Piandani$^4$, 
  Luca Pontisso$^6$, Davide Rossetti$^2$, Francesco Simula$^2$, Marco Sozzi$^5$ $^4$, 
  Laura Tosoratto$^2$, Piero Vicini$^2$}

\address{$^1$ INFN, Rome - Tor Vergata, Italy}
\address{$^2$ INFN, Rome - Sapienza, Italy}
\address{$^3$ CERN, Geneve, Switzerland}
\address{$^4$ INFN, Pisa, Italy}
\address{$^5$ Universiy, Pisa, Italy}
\address{$^6$ University, Rome, Italy}

\ead{alessandro.lonardo@roma1.infn.it}



\begin{abstract}

We implemented the \nanet FPGA-based \pcie Gen2 \gbe/\apelink NIC,
featuring GPUDirect RDMA capabilities and UDP protocol management
offloading. \nanet is able to receive a UDP input data stream from its
\gbe interface and redirect it, without any intermediate buffering or
CPU intervention, to the memory of a Fermi/Kepler GPU hosted on the
same \pcie bus, provided that the two devices share the same upstream
root complex. Synthetic benchmarks for latency and bandwidth are
presented. We describe how \nanet can be employed in the prototype of
the \mbox{GPU-based} RICH \mbox{low-level} trigger processor of the
NA62 CERN experiment, to implement the data link between the TEL62
readout boards and the low level trigger processor. Results for the
throughput and latency of the integrated system are presented and
discussed.
\end{abstract}


\section{Introduction}
The integration of GPUs in trigger and data acquisition systems is
currently being investigated in several HEP experiments.
At higher trigger levels, when the efficient \mbox{many-core}
parallelization of event reconstruction algorithms is possible, the
benefit of significantly reducing the number of the farm computing
nodes is evident~\cite{Clark:2010:GPUsAtlas, Rohr:2012:GPUsAlice}.
At lower levels, where tipically severe \realtime constraints are
present and custom hardware is used, the advantages of GPUs adoption
are less straightforward.
A pilot project within the CERN NA62
experiment\cite{Collazuol:2012zz} is investigating the usage of
GPUs in the central low level trigger processor, exploiting their
computing power to implement efficient, high throughput event selection
algorithms while retaining the \realtime requisites of the system.
One of the project preliminary results was that employing commodity
NICs and standard software stack caused data transfer over \gbe links
from readout boards to GPU memories to consume the largest part of the
time budget and was the main source of fluctuations in the overall
system response time.
In order to reduce data transfer latency and its fluctuations, we
envisioned the usage of the GPUDirect RDMA technology, injecting
readout data directly from the NIC into the GPU memories without any
intermediate buffering and the offloading of the network stack
protocol management from the CPU, avoiding OS jitter effects.
We implemented these two features in the \nanet \mbox{FPGA-based NIC}:
the first was inherited from the \apenetp 3D NIC
development~\cite{ammendola2012apenet+} while the second was realized
integrating an Open IP provided by the FPGA vendor.
\nanet NIC currently supports three \apelink 34~Gbps
channels~\cite{APEnetTwepp:2013} and one \gbe one; a 10\gbe version of
the design is under development.

After introducing the NA62 \mbox{multi-level} trigger system and
motivating the usage of GPUs in its \mbox{low-level} trigger
processor, we will provide a description of the \nanet architecture,
implementation and performances focusing on its usage as 1~\gbe NIC in
the case study of the \mbox{GPU-based} L0 trigger processor for the
RICH detector.

\section{The NA62 \mbox{multi-level} trigger system}

The NA62 experiment at CERN~\cite{Lamanna:2011zz} has the goal of
measuring the Branching Ratio of the \mbox{ultra-rare} decay of the
charged Kaon into a pion and a \mbox{neutrino-antineutrino} pair.
Due to the very high precision of theoretical prediction on this
Branching Ratio, a precise measurement at the level of 100 events
would be a stringent test of the Standard Model, also being this
Branching Ratio highly sensitive to any new physics particle.

Compared to first observations of this decay~\cite{Artamonov:2009sz},
the NA62 experiment aims at collecting more events ($\sim100$) with a
signal to background ratio 10:1, using a novel technique with
a \mbox{high-energy} (75~GeV) unseparated hadron beam decaying in
flight.
The experiment is currently in the final preparation stage, with the
first \mbox{data-taking} period foreseen for fall 2014.

The expected Standard Model Branching Ratio is $\simeq10^{-10}$,
requiring a very intense beam (main Kaon Branching Ratios is $\sim 10
\%$) and efficient background rejection.

The $\sim 10 MHz$ rate of particle decays reaching the detectors must
be reduced by a set of trigger levels down to a $\sim kHz$ rate.
The entire trigger chain works on the main digitized data
stream~\cite{Avanzini:2010zz}.
The first level (L0) is implemented in hardware (FPGAs) on the readout
boards and performs rather crude and simple cuts on the fastest
detectors, reducing the \mbox{high-rate} data stream by a factor 10 to
cope with the maximum design rate for event readout of 1~MHz.

Events passing L0 are transferred to the upper trigger levels (L1 and
L2) which are \mbox{software-implemented} on a commodity PC farm.

In the standard implementation, the readout boards FPGAs compute
simple trigger primitives on the fly, then \mbox{time-stamp} and send
them to a central processor for matching and trigger decision.
Thus, the maximum latency allowed for the synchronous L0 trigger is
related to the maximum data storage time available on the DAQ boards.
For NA62 this value is up to 1~ms, in principle allowing use of more
compute demanding implementations at this level, \ie the GPUs.

\subsection{The RICH detector low level (L0) trigger}

The RICH identifies pions and muons in the momentum range 15~$GeV/c$
to 35~$GeV/c$ with a $\mu$ suppression factor better than $10^{-2}$
with good time resolution.
\v{C}erenkov light is produced in a 18~m long, 3.7~m wide tube filled
with neon at atmospheric pressure.
The light is reflected by a composite mirror of 17~m focal length,
focused on two separated spots.
The two spots are equipped with $\sim 1000$~PMs of 1.8~cm in diameter
each.
After amplification and discrimination, the PM signal time is
digitized by high resolution TDCs.
A typical pion ring, for averaged accepted momentum, is identified
with $\sim 20$ firing PMs, as predicted by Monte Carlo and confirmed
with a \mbox{full-length} prototype~\cite{Angelucci:2010zz}.
Time resolution was measured to be better than 100~ps for all momenta
in the considered range.
Good time resolution and particle identification capability make this
detector ideal for use in the trigger system to build stringent
conditions.

\section{The GPU-based NA62 RICH detector L0 trigger} 

As a first example of GPU application in the NA62 trigger system we
studied the possibility to reconstruct rings in the RICH.
The center and the radius of the \v{C}erenkov rings in the detector
are related to particle angle/velocity.
This information can be employed at trigger level to increase the
purity and the rejection power for many triggers of interest.
The ring reconstruction could be useful both at L0 and L1.
In both cases, because of the high rate of 10 and 1~MHz respectively,
the computing power required is significant.
The GPUs can offers a simple solution of the problem.
The use of video cards in the L1 is straightforward: the GPU can act
as ``coprocessor'' to speed up the processing.
On the other hand, the L0 is a low latency synchronous level and
feasibility of GPU usage must be verified.
To test feasibility and performances, as a starting point we have
implemented five algorithms for single ring finding in a sparse matrix
of 1000 points (centered on the PMs in the RICH spot) with 20 firing
PMs (``hits'') on average.

\begin{figure}[t]
\begin{minipage}[t]{.49\textwidth}
\centering
  \includegraphics[trim=20mm 20mm 10mm 15mm,clip,width=\textwidth]{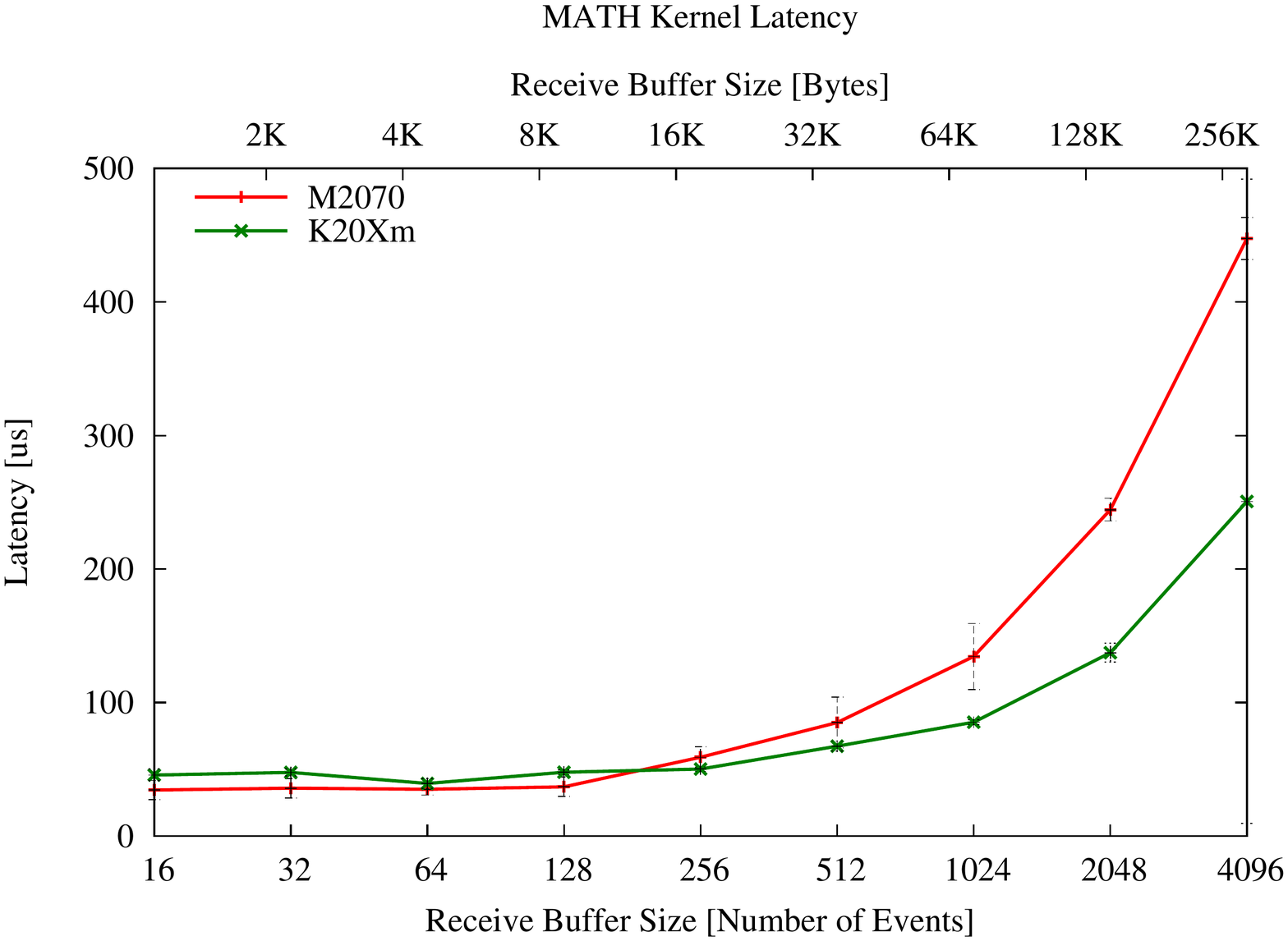}
  \caption{}
  \label{fig:lat_calc_MATH_M2070_K20x}
\end{minipage}
\quad
\begin{minipage}[t]{.49\textwidth}
  \centering 
  \includegraphics[trim=20mm 20mm 10mm 15mm,clip,width=\textwidth]{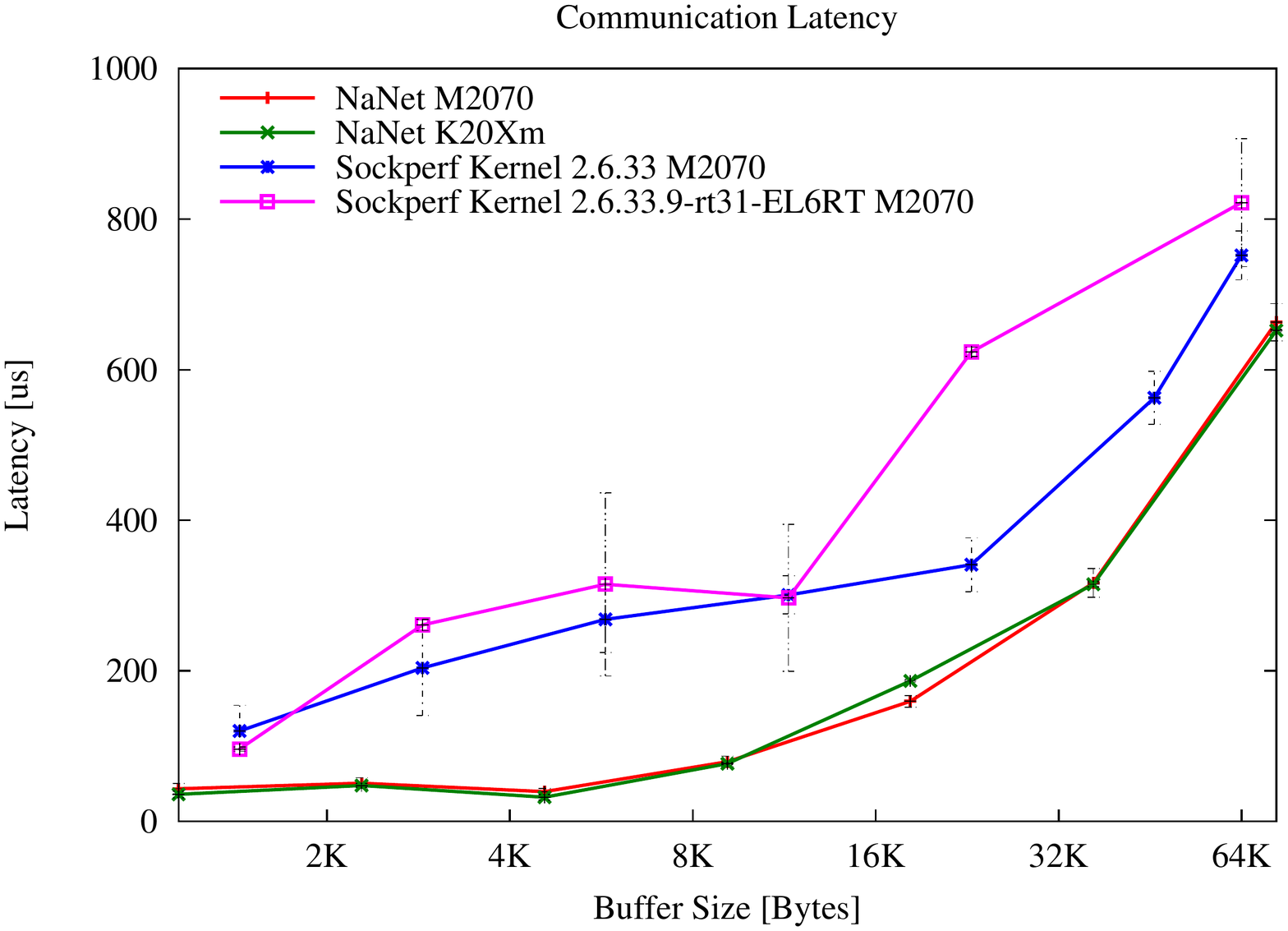}
  \caption{}
  \label{fig:lat_comm_vanilla_rt_nanet_M2070_K20x}
\end{minipage}
\vspace{-20pt}
\end{figure}

We tested these algorithms on TESLA C1060, C2050, M2070 and
K20~\cite{Collazuol:2012zz}.
In the following we focus on the fastest algorithm --- MATH --- where
the \mbox{least-squares} method is applied in a coordinate system in
which the problem can be analytically solved~\cite{Crawford:1983td}
with a linear inversion.
Processing times with input data and results in GPU memory, for the
MATH algorithm measured both on TESLA M2070 and K20Xm are plotted in
Fig.~\ref{fig:lat_calc_MATH_M2070_K20x}.
The contribution of processing to the overall system latency can be
kept under control due to the very small fluctuations in GPU kernel
execution times.

\subsection{Readout - L0 Trigger Processor Data Channel
  Implementation}
Data communication between the TEL62 readout boards and the L0 trigger
processor (L0TP) happens over multiple \gbe links using UDP streams.
The main requisite for the communication system comes from the request
for $<$1~ms and deterministic response latency of the L0TP:
communication latency and its fluctuations are to be kept under
control.
The requisite on bandwidth is 400$\div$700~MB/s, depending on the
final choice of the primitives data protocol which in turn depends on
the amount of preprocessing actually be implemented in the TEL62 FPGA.
So in the final system, 4$\div$6 \gbe links will be used to extract
primitives data from the readout board towards the L0TP.
We studied several options for the implementation of this multiple
\mbox{\gbe-based} data communication system, benchmarking any of them
for a single \gbe channel in order to collect indications for the
design of the \mbox{full-fledged} system.
A first result was that any solution matched the bandwidth
specification for a \gbe link at significant buffer sizes, so we
concentrated on measuring communication latency and, most important in
the context of the design of a \realtime communication system, latency
fluctuations.
To perform benchmarks we used two different hardware platforms:
\begin{compactitem}
\item a Supermicro SuperServer 6016GT-TF with X8DTG-DF motherboard
  (Intel 5520-Tylersburg chipset), dual Intel Xeon X5570 @2.93~GHz
  CPU, Intel 82576 \gbe and \nvidia Fermi M2070 GPU (from here on
  M2070 system)
\item a Supermicro SuperServer 7047GR-TPRF with X9DRG-QF motherboard
  (Intel C602-Patsburg chipset), dual Intel Xeon E5-2609 @2,40~Ghz
  CPU, Intel i350 \gbe and \nvidia Fermi K20Xm GPU (from here on K20Xm
  system).
\end{compactitem}

First option considered was a standard Linux installation (CentOS 6.3,
Kernel 2.6.33) with integrated \gbe interface in the M2070 system; to
measure latencies we used the network benchmarking
utility \emph{sockperf}\cite{sockperf}.
Results are shown in
Fig.~\ref{fig:lat_comm_vanilla_rt_nanet_M2070_K20x}; at
lower buffer sizes latencies are higher than desirable but main
drawback of this setup is the great latency variability.

Next option in the attempt of reducing latency fluctuations was trying
a \realtime kernel on the M2070 system.
A great effort has been recently done by OS developers in improving RT
features in kernels: predictability in response times, reduced
jitters, $\mu$s accuracy and improved time granularity.
In Fig.~\ref{fig:lat_comm_vanilla_rt_nanet_M2070_K20x} results
obtained with a \texttt{2.6.33.9-rt31-EL6RT} kernel are plotted;
CPUspeed and IRQbalance daemons were stopped and Interrupt moderation
was disabled to avoid other possible sources of latency fluctuations.
This approach was successful in minimizing fluctuations on latency but
increased the latency values up to an incompatible level with the L0TP
1~ms time budget.

Another considered option was usage of \texttt{\emph{PF
    RING}}~\cite{PFRING}, which is a framework for accelerating packet
capture implementing a \mbox{memory-mapped} buffer allocated at socket
creation, \ie where incoming packets are copied.
\texttt{PF RING} can use either standard drivers or \mbox{\texttt{PF
    RING}-aware} drivers and works with \mbox{off-the-shelf} \gbe
NICs.
Promising results obtained using this approach are reported and
discussed in~\cite{CHEP2013:GAP}.

Finally, to tackle the \realtime requirement of the \mbox{GPU-based}
L0TP, we considered reusing the GPUDirect RDMA technology that we
already implemented in the \apenetp project for \mbox{3D-torus}
network card.
This led to the design and implementation of the \nanet
\mbox{FPGA-based} NIC featuring, besides GPUDirect RDMA capability, a
UDP offloading engine.
Latency benchmarks obtained using \nanet both in the M2070 and the
K20Xm system are shown in
~\ref{fig:lat_comm_vanilla_rt_nanet_M2070_K20x}.
Latency and its variability are significantly reduced when compared to
other benchmarked solutions.
In the following sections we describe the internal architecture of
\nanet and report a performance analysis for it and the
\mbox{GPU-based} RICH L0TP using \nanet as a communication channel
from the readout boards.
 
\section{\nanet}
\label{sec:nanet}
\nanet is an \apenetp rehaul for \mbox{real-time} data acquisition
able to inject directly data from the NIC into the CPU/GPU memory with
no intermediate buffering, reusing the \apenetp GPUDirect RDMA
implementation.
Moreover, it adds a network stack protocol management offloading
engine to the logic to avoid OS jitter effects.

\begin{figure}[t]
  \begin{minipage}[t]{.49\textwidth}
    \centering
    \includegraphics[trim=60mm 25mm 60mm 30mm, clip,width=0.8\textwidth]{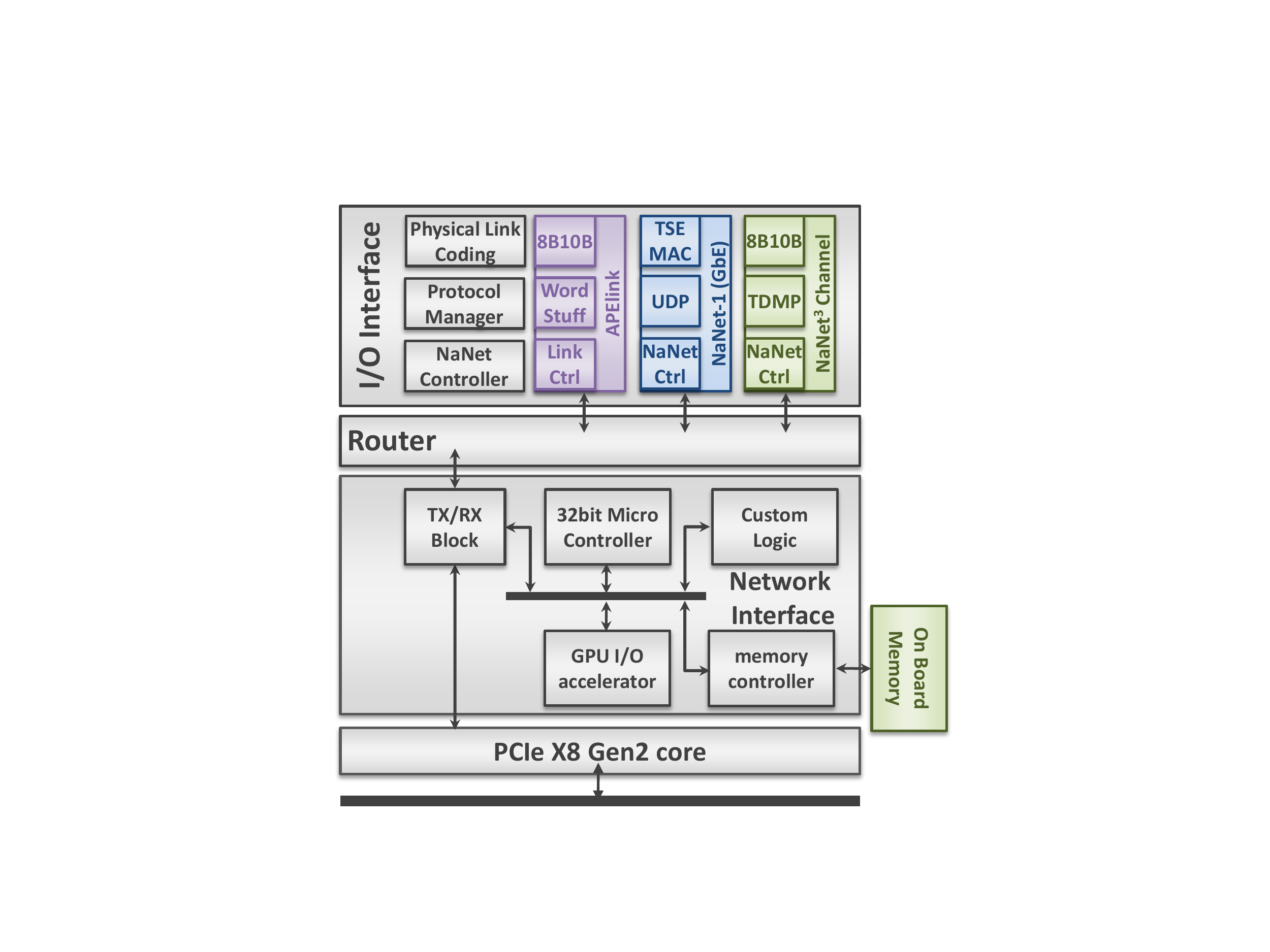}
    \caption{\nanet features a customizable I/O interface to implement
      a low latency, \realtime NIC for hybrid CPU/GPU systems.}
    \label{fig:NaNet}
  \end{minipage}
  \quad
  \begin{minipage}[t]{.49\textwidth}
    \centering 
    \includegraphics[width=\textwidth]{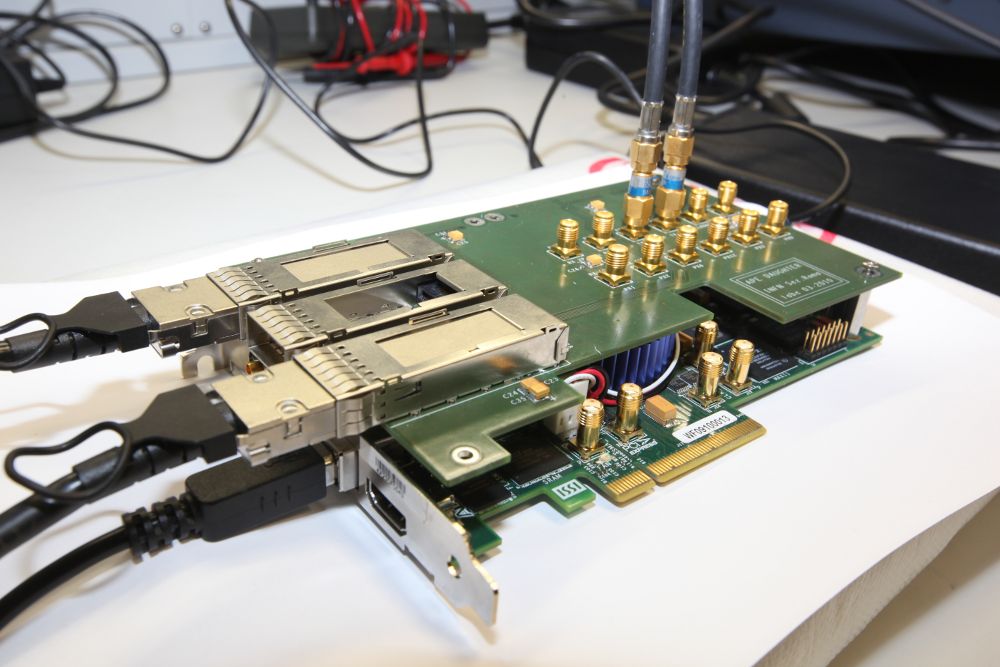}
    \caption{\nanetone implemented on an Altera Stratix IV coupled
      with a custom mezzanine card sporting 3 \apelink channels.}
    \label{fig:board}
  \end{minipage}
  \vspace{-20pt}
\end{figure}

\nanet design supports a configurable number and kind of I/O channels
(see figure~\ref{fig:NaNet}); incoming data streams are processed by a
Physical Link Coding block feeding the Data Protocol Manager that in
turns extracts the payload data.
These payload data are encapsulated in the \apenetp data packet
protocol by the \nanet Controller and sent to the \apenetp Network
Interface, taking care of their delivery to the destination memory.

\subsection {\nanetone Hardware Architecture} 
\label{sec:nanetone}

The \nanetone is a \pcie Gen2 x8 NIC featuring a standard \gbe
interface able to directly inject an UDP data stream into the memory
of a Fermi- or \mbox{Kepler-class} \nvidia GPU leveraging on GPUDirect
RDMA capabilities, implemented on a Stratix IV GX FPGA Dev Kit.
Moreover, it provides 3 \mbox{bi-directional} \apelink channels, with
the addition of a custom mezzanine equipped with 3 QSFP+ connectors.

The \gbe trasmission is designed following the general I/O interface
architecture pointed out in figure~\ref{fig:NaNet}.
Physical Link Coding is Altera Triple Speed Ethernet Megacore (TSE
MAC), providing 10/100/1000~Mbps Ethernet IP modules.
The \textit{UDP offloader} collects data coming from the TSE MAC,
extracting UDP packets payload and providing a \mbox{32-bit} wide
channel achieving 6.4~Gbps, discharging the \nios from the data
protocol management.
Finally, the \textit{NaNet CTRL} is the hardware module in charge of
encapsulating the UDP data in the proprietary \apenetp protocol,
parallelizing incoming \mbox{32-bit} data words into \mbox{128-bit}
\apenetp ones.

The \textit{Network Interface}, the packet injection/processing logic
providing hardware support for Remote Direct Memory Access (RDMA)
protocol for CPU and GPU and the \textit{Router} with I/O channels
multiplexing tasks are inherited from \apenetp.
\subsection {\nanetone Software Stack}

Software components for \nanetone operation are needed both on the x86
host and on the \nios \mbox{FPGA-embedded} $\mu$controller.
On the x86 host, a GNU/Linux kernel driver and an application library
are present.
The application library provides an API mainly for \texttt{open/close}
device operations, registration (\ie allocation, pinning and returning
of virtual addresses of buffers to the application) and deregistration
of circular lists of persistent receiving buffers (CLOPs) in GPU
and/or host memory and signalling of receive events on these
registered buffers to the application (\eg to invoke a GPU kernel to
process data just received in GPU memory).
On the $\mu$controller, a single process application is in charge of
device configuration, generation of the destination virtual address
inside the CLOP for incoming packets payload and virtual to physical
memory address translation performed before the \pcie DMA transaction
to the destination buffer takes place.

\section{Performance Analysis}
\begin{figure}[t]
  \begin{minipage}[t]{.49\textwidth}
    \centering
    \includegraphics[trim=20mm 20mm 10mm 20mm,clip,width=\textwidth]{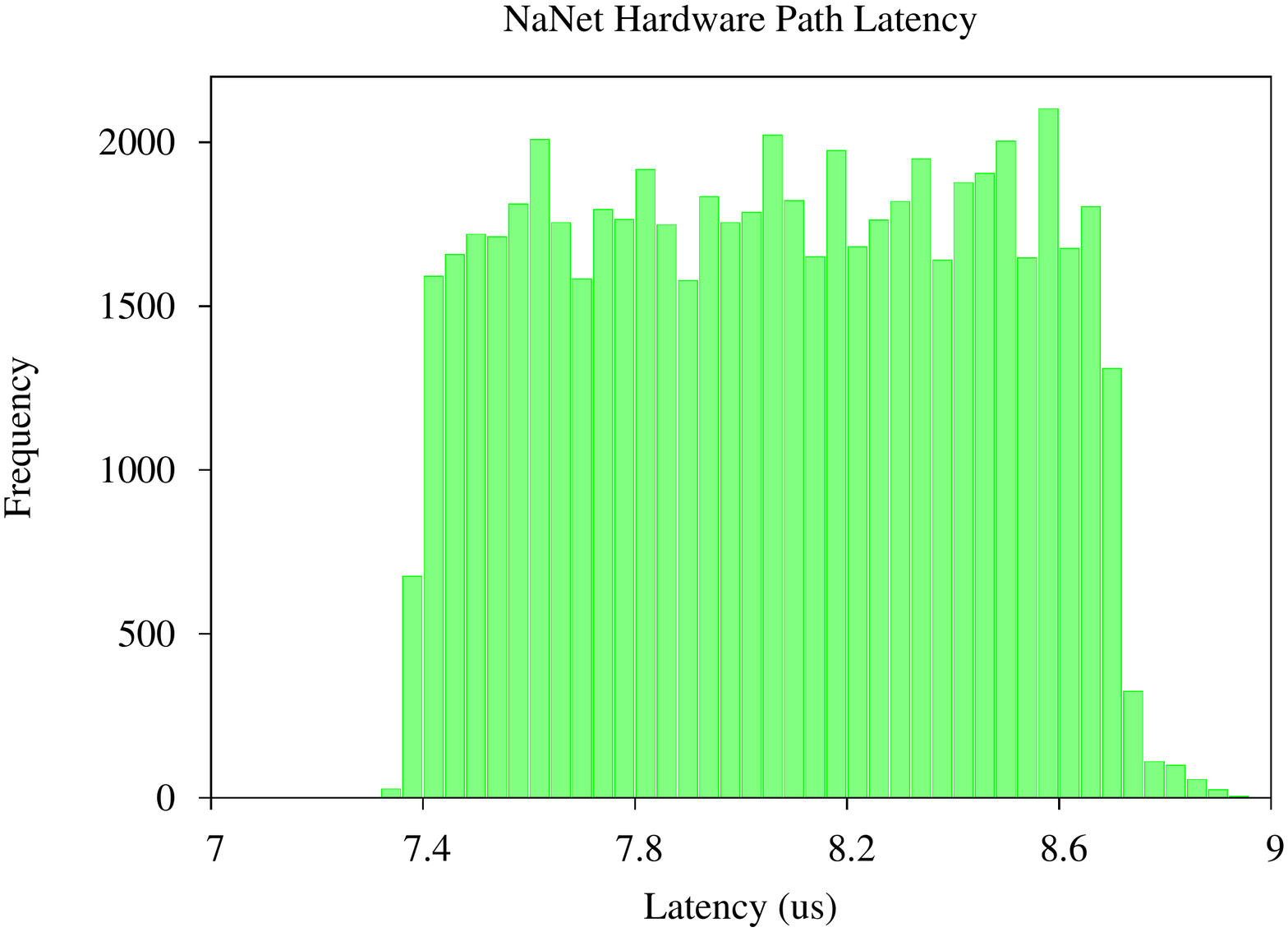}
    \caption {}
    \label{fig:lat_hw_path}
  \end{minipage}
  \quad
  \begin{minipage}[t]{.49\textwidth}
    \centering 
    \includegraphics[width=.8\textwidth]{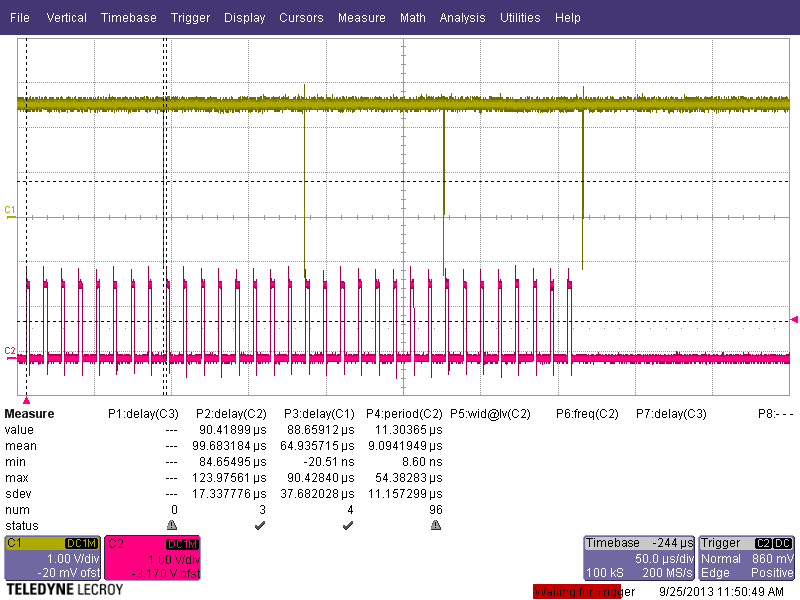}
    \caption{}
    \label{fig:lat_tel62_32pkt_clop8}
  \end{minipage}
  \vspace{-10pt}
\end{figure}
We measured \nanetone latency and bandwidth using different methods,
then we tested it integrated in a simulated \mbox{GPU-based} RICH L0
trigger processor, measuring performances (latency and throughput) of
the overall system.


Latency of \nanetone NIC was benchmarked using several methods.

Firstly, we instrumented the FPGA logic with a dedicated hardware path
traversal latency measurement system able to add a ``profiling''
footer to the packet payload, storing up to 4 cycle counters values
recorded at different packet processing stages.
We were thus able to characterize the latency associated to processing
in relevant \nanetone subsystems, namely the {UDP offloader}, the
\nios $\mu$controller and the {Tx} block in the {Network Interface}.
In Fig.~\ref{fig:lat_hw_path} a histogram is plotted with hardware
processing path traversal latency inside \nanetone: values show an
appreciable variability, due to the \nios $\mu$controller performing
address generation and virtual to physical translation tasks.
This clearly indicates the need for a redesign, implementing dedicated
FPGA logic blocks performing these two tasks.

A second method was using one of the host \gbe ports to send UDP
packets according to the NA62 RICH readout data protocol to the
\nanetone \gbe interface: using the x86 TSC register as a common
reference time, it was possible in a single process test application
to measure latency as time difference between when a received buffer
is signalled to the application and the moment before the first UDP
packet of a bunch (needed to fill the receive buffer) is sent through
the host \gbe port.
Within this measurement setup (``system loopback''), the latency of
the send process is also taken into account.
Measurements in Fig.~\ref{fig:lat_comm_vanilla_rt_nanet_M2070_K20x}
were taken using this method; UDP packets with a payload size of
1168~B (16 events) were sent to a GPU memory receiving buffer of size
variable between 1 and 64 UDP packet payload sizes.

Connecting a TEL62 readout board sending \mbox{Monte Carlo-generated}
events stored onto the FPGA through one of its \gbe ports to a
\nanetone board, we were able, besides testing the integration of our
NIC in the working environment, to perform oscilloscope latency
measurements as depicted in Fig.~\ref{fig:lat_tel62_32pkt_clop8}: a
bunch of 32 UDP packets is sent from the TEL62 readout board (red
signal) and 4 \pcie completion (yellow signal) show the end of the
\pcie DMA write transaction towards the GPU memory buffers, each sized
8 times the UDP packet payload size.

\begin{figure}[t]
  \begin{minipage}[t]{.49\textwidth}
    \centering \includegraphics[trim=20mm 20mm 10mm
      15mm,clip,width=\textwidth]{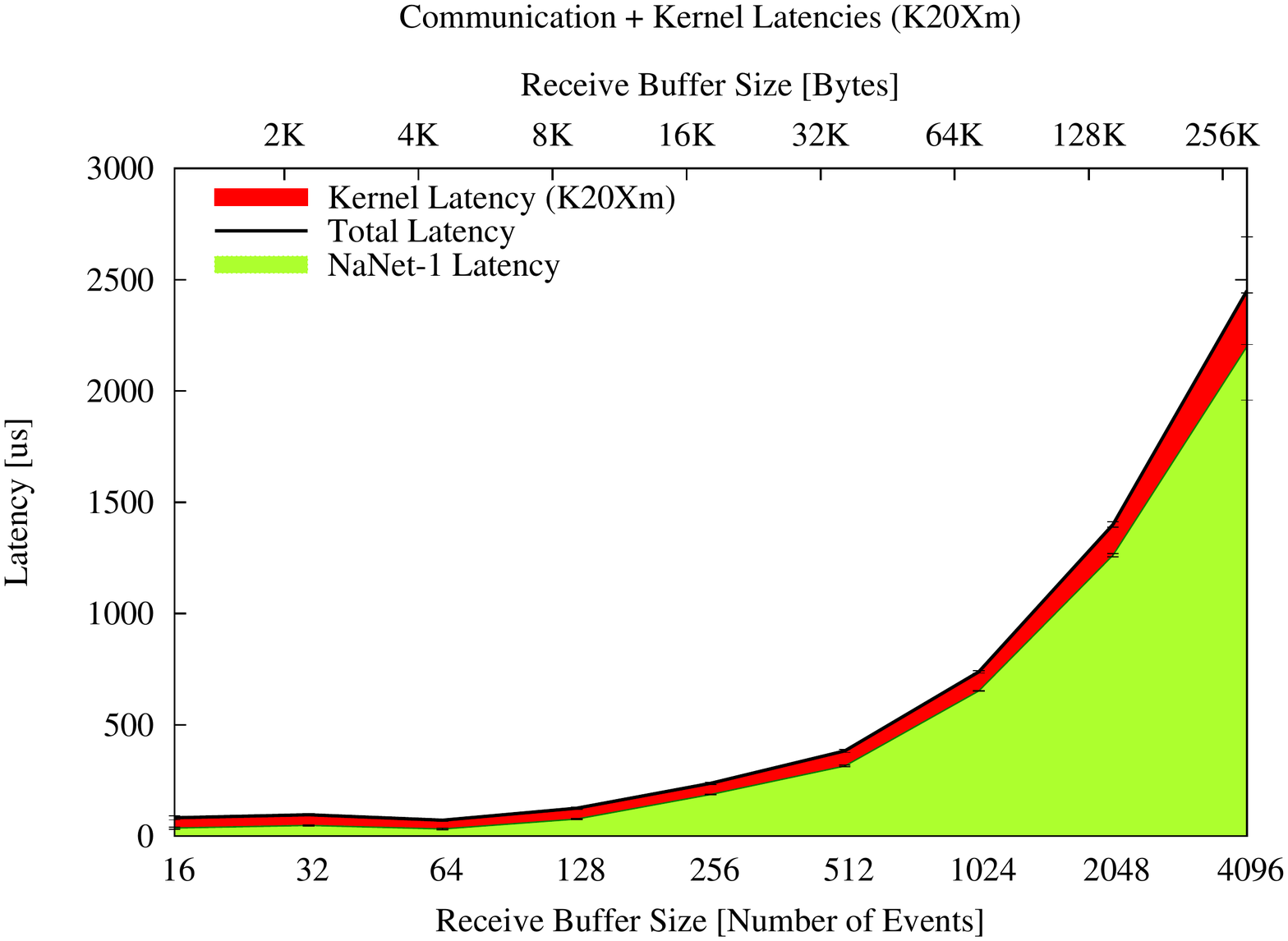}
    \caption {}
    \label{fig:ap3iron2}
  \end{minipage}
  \quad
  \begin{minipage}[t]{.49\textwidth}
    \centering 
    \includegraphics[trim=20mm 20mm 10mm 15mm,clip,width=\textwidth]{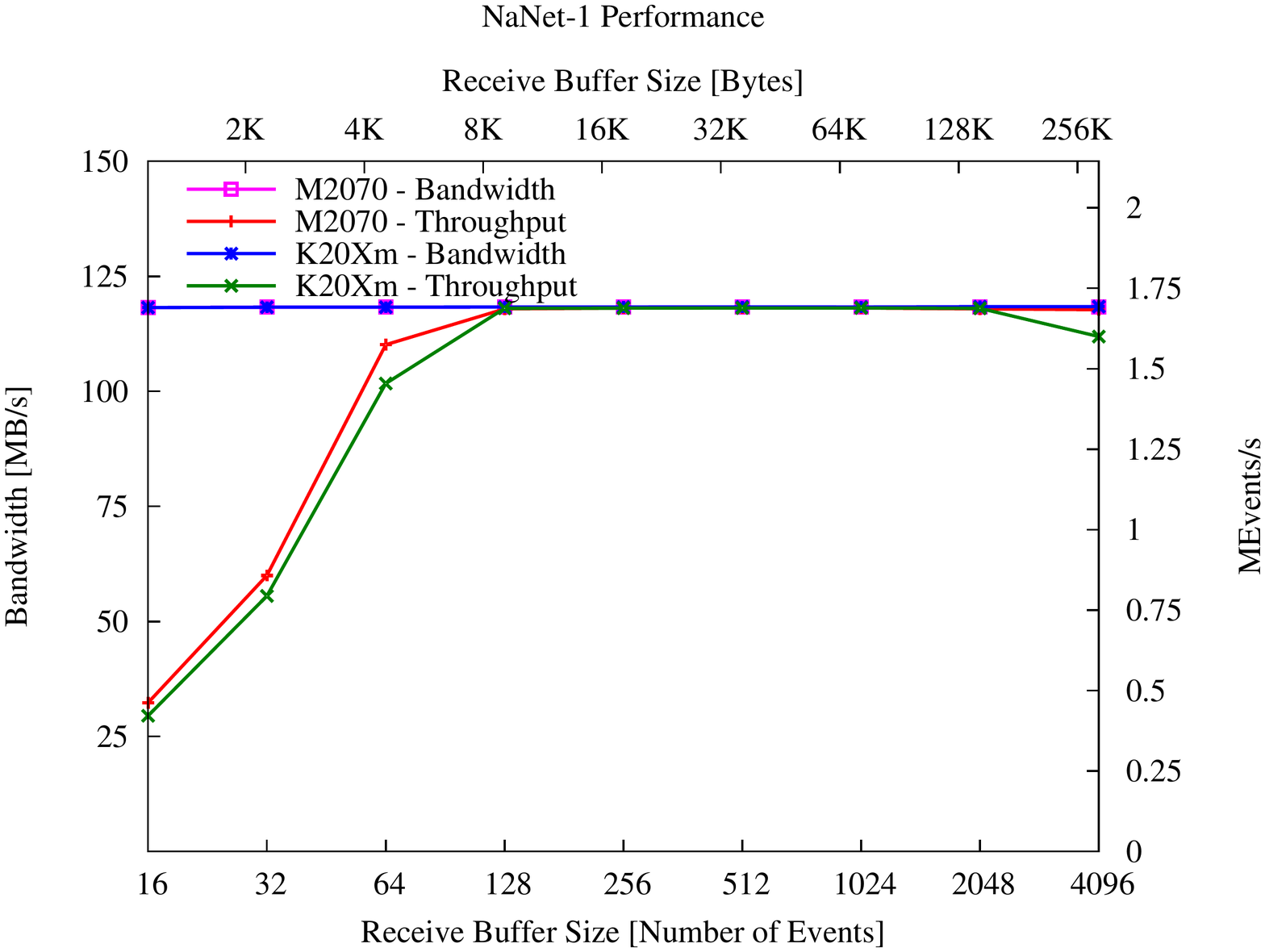}
    \caption{}
    \label{fig:throughput}
  \end{minipage}
  \vspace{-20pt}

\end{figure}

As anticipated, bandwidth measurement was also performed, both for the
M2070 and the K20Xm system: results are in Fig.~\ref{fig:throughput}.

A \mbox{GPU-based} L0TP setup scaled down in bandwidth was reproduced
by using a system loopback configuration, with the host system
simulating the TEL62 UDP traffic through one of its \gbe ports towards
a \nanetone NIC redirecting incoming data stream towards a GPU memory
circular list of receive buffers; once received, such buffers are
consumed by a CUDA Kernel implementing the MATH \mbox{ring-finding}
algorithm.
Communication and kernel processing tasks were serialized in order to
perform the measure; these are the results for the K20Xm system in
Fig.~\ref{fig:ap3iron2}, representing a \mbox{worst-case} situation.
During normal operation, this serialization constraint can be relaxed, 
and kernel processing task overlaps with data communication.
Actually this is what has been done to measure system throughput,
results are shown in Fig.~\ref{fig:throughput}.
Combining the two results, we see that using GPU receive buffer sizes
ranging from 128 to 1024 events allow the system to remain within the
1~ms time budget while keeping a $\sim1.7$~MEvents/s throughput.

\section{Conclusions and Future Work}
Our \nanet design proved to be efficient in performing \realtime data
communication between the NA62 RICH readout system and the GPU-based L0
trigger processor over a single GbE link.
These encouraging results are corroborated by benchmarks carried on
using one \apelink 34~Gbps channel supported by \nanetone ~\cite{NanetTwepp:2013}.
To cope with the full system bandwidth requirement we started developing a \nanet design supporting dual \tengbe on SFP+ ports.

\section*{Acknowledgment}
This work was partially supported by the EU Framework Programme 7 project EURETILE under grant number 247846.
G. Lamanna, F. Pantaleo, R. Piandani and M. Sozzi thank the GAP
project, partially supported by MIUR under grant RBFR12JF2Z ``Futuro
in ricerca 2012''; R. Ammendola was supported by MIUR through INFN SUMA project.

\end{document}